\documentclass[12pt,preprint]{aastex}

\begin{document}

\title{The Antennae Ultraluminous X-Ray Source, X-37, \\ Is A Background Quasar}

\author{D.M. Clark \altaffilmark{1}, M.H. Christopher
\altaffilmark{2}, S.S. Eikenberry \altaffilmark{1}, B.R. Brandl
\altaffilmark{3}, J.C. Wilson \altaffilmark{4}, J.C. Carson
\altaffilmark{5}, C.P. Henderson \altaffilmark{6}, T. L. Hayward
\altaffilmark{7}, D.J.  Barry \altaffilmark{6}, A.F. Ptak
\altaffilmark{8}, E.J.M. Colbert \altaffilmark{8}}

\altaffiltext{1}{Department of Astronomy, University of Florida,
Gainesville, FL 32611; dmclark@astro.ufl.edu}
\altaffiltext{2} {Caltech, Department of Astronomy, 105-24, Pasadena,
CA 91125.}
\altaffiltext{3}{Leiden Observatory, 2300 RA Leiden,
Netherlands.}  
\altaffiltext{4}{Department of Astronomy, P.O Box 3818, University of
Virginia, Charlottesville, VA 22903.}
\altaffiltext{5}{JPL, Earth \& Space Science, Pasadena, CA 91109.}
\altaffiltext{6}{Astronomy Department, Cornell University, Ithaca, NY
14853.}
\altaffiltext{7}{Gemini Observatory, AURA/Casilla 603, La Serena,
Chile}
\altaffiltext{8}{Department of Physics and Astronomy, Johns Hopkins
University, 3400 North Charles St., Baltimore, MD 21218.}

\begin{abstract}

In this paper we report that a bright, X-ray source in the Antennae
galaxies (NGC 4038/9), previously identified as an ultra-luminous
X-ray source, is in fact a background quasar.  We identify an isolated
infrared and optical counterpart within $0.3\pm0\farcs5$ the X-ray
source X-37.  After acquiring an optical spectrum of its counterpart,
we use the narrow [OIII] and broad H$\alpha$ emission lines to
identify X-37 as a quasar at a redshift of z=0.26. Through a $U$, $V$,
and $K_s$ photometric analysis, we demonstrate that most of the
observable light along this line of sight is from the quasar.  We
discuss the implications of this discovery and the importance of
acquiring spectra for optical and IR counterparts to ULXs.

\end{abstract}


\section{Introduction}

Ultraluminous X-ray sources (ULX) are typically defined as point
sources with X-ray luminosities $> 10^{39}$ ergs s$^{-1}$.  {\it
Einstein} \citep{lon83,hel84,fab89} and {\it ROSAT}
\citep{rob00,col02} observations revealed many of these objects in
nearby galaxies. Recent {\it Chandra} observations indicate the
Antennae (NGC 4038/4039) contain 9 ULXs \citep[assuming here and hence
forth $H_0$ = 75 km s$^{-1}$ Mpc$^{-1}$]{zez02a}, the largest number
discovered so far in a single galaxy \citep{fab03}.

One X-ray source in the Antennae that has received considerable
attention in the recent literature is X-37, as designated by
\citet{zez02a}.  At the distance to these interacting galaxies (19.3
Mpc) it would have an X-ray luminosity, $L_x$ = 4.5$\times$$10^{39}
$ergs s$^{-1}$, making it a ULX. A previous attempt to match X-ray
positions directly to {\it HST} positions indicated that this object
has a significant ($> 1\farcs0$) offset from a nearby optical source
\citep{zez02b}.  This spawned discussions of whether this is a runaway
X-ray binary escaping from its parent cluster
\citep{zez02b,mil04a,fab04}.  But, matching X-ray positions directly
to the optical is difficult due to the crowded {\it HST} field and
image rotation.

In \citet{cla05} (henceforth Paper 1) we demonstrated that the
infrared (IR) is a powerful method for finding counterparts to
X-ray sources.  This wavelength regime has similar dust penetrating
properties to X-rays, facilitating counterpart identification.  In a
subsequent paper we match X-ray and optical {\it HST} positions with
high precision by using the IR frame tie as a go-between (Clark, et
al. 2005, in preparation; henceforth Paper 2).  Here we use X-37 to
demonstrate the success of our method and show the importance of
obtaining follow-up spectra of counterparts to ULX candidates. \S2
describes our spectroscopic observations and briefly discusses our
analysis of the IR and optical images.  We summarize our results and
present the implications of our discovery in \S3.

\section{Observations and Data Analysis}

\subsection{Infrared and Optical Images}

Our analysis of the counterpart to X-37 is based in part on IR $J$
(1.25 $\mu$m) and $K_s$ (2.15 $\mu$m) band images of the Antennae.
These images were first analyzed by \citet{bra05}, who reported the
details of the data reduction.  In summary, we acquired 20 minute
total exposures in each filter using the Wide-field InfraRed Camera
(WIRC) on the Palomar 5-m telescope during the night of March 22nd,
2002.

In Paper 1, we made a frame tie between the IR and X-ray using IR
counterparts to compensate for the poor astrometric accuracy of {\it
Chandra} ($\sim1\farcs5$).  We matched 7 IR sources from the WIRC
images with {\it Chandra} X-ray point sources.  Using a least squares
fit of a linear matching function we tied {\it Chandra} R.A. and
Dec. to WIRC x,y pixel positions.  The RMS positional uncertainty is
$\sim0\farcs5$.  

With the astrometric frame tie in place, we identified a bright, $K_s$
= 16.2 mag source $0.3\pm0\farcs5$ from the X-ray position of X-37
(see Figure 1).  Note that this contradicts previous reports of a
significant offset between X-37 and the star cluster candidate, and
independently eliminates any need for the hypothesis that this
X-ray source is a runaway X-ray binary (Zezas \& Fabbiano 2002; see
also Miller et al.  2004a; Fabbiano et al. 2003).

We also base this study on work done in Paper 2 where we studied
optical counterparts to X-ray sources in the Antennae.  We obtained
archival {\it HST} images \citep{whi99} in the following filters:
F336W($U$), F439W($B$), F555W($V$), and F814($I$).  Tying {\it
Chandra} X-ray coordinates to {\it HST} positions is nontrivial due to
image rotation and the crowded field in the {\it HST} images.
Instead, we used our excellent frame tie between the IR and X-ray as
an intermediate step.  We used the same basic procedure as that used
in \citet{cla05}.  Using this astrometric frame tie, we found an
optical source $0.6\pm0\farcs6$ from the X-ray position of X-37
(Figure 1) --- again, an insignificant offset.

\subsection{Spectroscopy}

We acquired an optical spectrum of X-37 on March 7, 2003, using the
Low Resolution Imaging Spectrograph \citep[LRIS]{oke95} on Keck I. We
observed this source in two separate slitmasks with a 400 lines
mm$^{-1}$ grating and blaze wavelength of 8500 \AA, yielding a total
wavelength coverage on the red side of 5600-9400 \AA.  Seeing
throughout the observations was $\sim0\farcs8$.  We observed X-37 for
a total of 6000 seconds between the two masks with typical single
exposure times of 600 seconds.

We reduced the spectrum with a combination of standard IRAF procedures
and IDL code using the NeAr arc-lamp spectrum for wavelength
calibration.  We used the standard star Feige 67 to flux calibrate the
spectrum.

We extracted spectra of the X-37 counterpart from each individual
exposure in a $1\farcs8$ (approximately 2 FWHM) region centered on the
emission peak. A background spectrum extracted from a region
$2\arcsec$ removed from the counterpart was sufficient for removing
the night sky lines.  We point out the blended lines, H$\alpha$/[NII]
at $\sim6600$ \AA$ $ and the [SII] pair at $\sim6750$ \AA, observed at
the redshift of the Antennae (Figure 2).  These spectral features are
at approximately the same levels in the on-source and background
spectra, suggesting that they are likely diffuse emission from the
Antennae and not associated with X-37.

Noting the broad emission lines characteristic of a quasar we looked
for other spectral features to confirm this classification.  The
realization that the broad line at 8296 \AA$ $ was red-shifted
H$\alpha$ prompted a search for additional lines generally associated with
quasars.  We noted a plethora of such identifiers including the
non-restframe wavelengths of H$\beta$ at 6148 \AA, [OI] at 7965 \AA,
[NII] at 8320 \AA, and [SII] at 8510 \AA.  This corroborated our
conclusion that X-37 is a quasar.

Using the observed, narrow [OIII] lines at 6270.45 \AA$ $ and 6330.94
\AA, we measured a redshift of z=0.26.  At the redshift to this
quasar, X-37 would now have an X-ray luminosity of $L_x = 1.4 \times
10^{43}$ ergs s$^{-1}$.  As we show in Table 1, this luminosity is
typical of the X-ray luminosity for quasars at a similar distance
\citep{sch96}.

\subsection{Photometry}

To investigate how much of the continuum flux could be due to the
quasar, we compared the photometric properties of this source to that
of other quasars at X-37's redshift of z=0.26.  We chose $U$ and
$V$-band photometry from the {\it HST} images and $K_s$-band
photometry from the WIRC images.  This choice in filters covers the
full wavelength range used in our studies of the Antennae (Clark, et
al. 2005, in preparation).  The full-width at half-maximum (FWHM) in
$U$, $V$, and $K_s$ are $0\farcs24$, $0\farcs18$ and $0\farcs9$,
respectively.  Defining the photometric aperture as $\sim3\sigma$ of
the Gaussian point spread function (PSF) width, the aperture radii are
3-pixels in $U$, 2.5-pixels in $V$, and 5-pixels in $K_s$.

Subtracting sky background flux is especially difficult in the
Antennae considering the bright, crowded background in these galaxies.
To compensate for this difficulty, we chose to measure the mean and
median sky background flux in two separate annuli between $\sim6 -
10\sigma$ of the PSF.  Being an isolated source, we did not have to
compensate for crowded conditions in the field around the X-37
counterpart.  Multiplying the mean and median sky background by the
area of the central aperture and then subtracting this from the
central aperture yielded four separate flux measurements for the
source.  We then averaged these four values to give us the source
flux.  When computing errors, we considered variations in the sky
background, $\sigma_{sky}$, and Poison noise, $\sigma_{adu}$.  To
calculate $\sigma_{sky}$ we found the standard deviation of the four
flux values for each source.  Dividing the mean source flux by the
instruments gain and then taking the square root of these, we found
$\sigma_{adu}$.  The known gain for WIRC is 2 e$^{-1}$/DN
\citep{wil03}, while \citet{whi99} use a gain of 7 e$^{-1}$/DN for
{\it HST}.  Both $\sigma_{sky}$ and $\sigma_{adu}$ were added in
quadrature to yield a total error in flux, $\sigma_{flux}$.

We used a bright, 2MASS star to convert $K_s$ flux into a magnitude.
We derived {\it HST} magnitudes using zero points listed in Table 28.1
of the {\it HST} Data Handbook \citep{voi97}.  Applying color
transformations defined in \citet{hol95}, we converted {\it HST}
magnitudes to Johnson $U$ and $V$.  We expressed errors in magnitude,
$\sigma_m$, as $\sigma_{flux}$ divided by the mean flux.  In the case
of $U$ and $V$, $\sigma_m$ consists of the additional errors in the
zeropoint and color transformations, all of which were added in
quadrature.

\subsubsection{Absolute Magnitudes}

We next converted these photometric measurements to
reddening-corrected absolute magnitudes.  At the redshift of X-37, the
distance modulus is 40.1 mag.  We derived reddening in $V$, $A_V$ = 1.3
mag, using the X-ray derived column density, $N_H$, listed in Table 5
of \citet{zez02a} and the relationship $A_V$ = $5\times10^{-22}$
$cm^{-2}$ $N_H$.  We used the $N_H$ value provided by the power-law
spectral fits of \citet{zez02a}.  These authors computed $N_H$
assuming the absorption is at zero redshift.  This should not be a
problem for us since the broad absorption lines observed in the X-37
spectrum suggest there is little extinction along the line of sight to
it.  

To find reddening in other filters, we used the extinction law
defined in \citet[CCM]{car89}.  We also computed separate reddenings
in each filter using the Rieke-Lebofsky (RL) Law \citep{rie85}.  We
then expressed errors in reddening, $\sigma_A$, as the difference in
$A_U$, $A_V$ and $A_{K_s}$ derived from each law.  Adding $\sigma_A$
and $\sigma_m$ in quadrature, we computed a total error in absolute
magnitude.

In Table 1, we compare the absolute magnitudes, $M_U$, $M_V$, and
$M_{K_s}$, of the X-37 counterpart to typical magnitudes for quasars
at a similar distance.  We obtained $M_U$ and $M_V$ from the Sloan
Digital Sky Survey (SDSS) quasar catalog \citep{sch03} for 33 quasars
at a redshift of 0.26.  Using color transformations listed in
\citet{fuk96}, we converted SDSS magnitudes to the Johnson photometric
system.  We used the catalog described in \citet{bar01} to derive
$M_{K_s}$ for 17 quasars at a redshift of 0.26.  Considering the
absolute magnitude of X-37 in the $U$, $V$ and $K_s$ bands falls within
the range of catalog quasar magnitudes, X-37 has the luminosity of a
typical quasar.

\section{Discussion}

The identification of X-37 with a background quasar further
demonstrates the importance of spectroscopic followup for the study of
ULX sources.  While the overabundance of such sources near galaxies
clearly demonstrates that a physical connection does exist for many
ULX sources as a population \citep[e.g.][]{col02}, the strong
possibility of background quasar contamination makes such
identification for any particular source perilous \citep[see
also][]{gut05}.

A specific example involving X-37 is the recent work of
\citet{mil04b}.  In their Figure 2, \citet{mil04b} compare the black
hole mass to inner disk temperature for both ULX sources and
``standard" stellar-mass black holes, finding that the ULX sources
have cooler disk temperatures.  They use this to conclude that ULX are
likely to be powered by accretion onto Intermediate-Mass Black Holes
(IMBH).  While some of the sources in this diagram are almost
certainly ULXs (i.e. NGC 1313 X-1 and NGC 1313 X-2), and thus possible
IMBH, it is somewhat surprising to note that the ``ULX" X-37 is
grouped with these sources in the diagram -- despite the fact that its
actual black hole mass is at least 4 orders of magnitude higher than
assumed by \citet{mil04b}.  Thus, X-37 illustrates the importance of
follow-up spectroscopic confirmation of ULXs, and sounds a cautionary
note regarding conclusions based on observations without such confirmation.

\acknowledgments

The authors are especially grateful to Fred Hamann and Vicki
Sarajedini for examining our spectra and helping identify this unusual
source as a quasar.  We also thank J. Houck for his support of the
WIRC instrument project.  WIRC was made possible by support from NSF
(NSF-AST0328522), the Norris Foundation, and Cornell University.  SSE
and DMC are supported in part by an NSF CAREER award (NSF-9983830).

\vfill \eject

\clearpage

\begin{deluxetable}{lcc}
\tablecaption{Absolute Magnitude Comparison to Quasars at $z\approx0.26$}
\tablewidth{0pt}
\tablehead{
Filter & Observed Magnitudes \tablenotemark{1} & Typical Catalog Range}
\startdata
$U$ & -20.9$\pm$0.2 & -23.9 -- -20.8 \\
$V$ & -20.4$\pm$0.2 & -23.4 -- -21.0 \\
$K_s$ & -24.1$\pm$0.1 & -25.6 -- -23.8 \\
\tableline
log($L_x$) & 43.1 & 43.7 -- 44.1 \\
\enddata
\tablenotetext{1}{Absolute magnitudes include uncertainties in each
  value.  Catalog values were taken from separate sources: $U$ and $V$
  from \citet{sch03}, $K_s$ from \citet{bar01}, and $L_x$ from \citet{sch96}.}
\end{deluxetable}

\clearpage

\begin{figure}
\plottwo{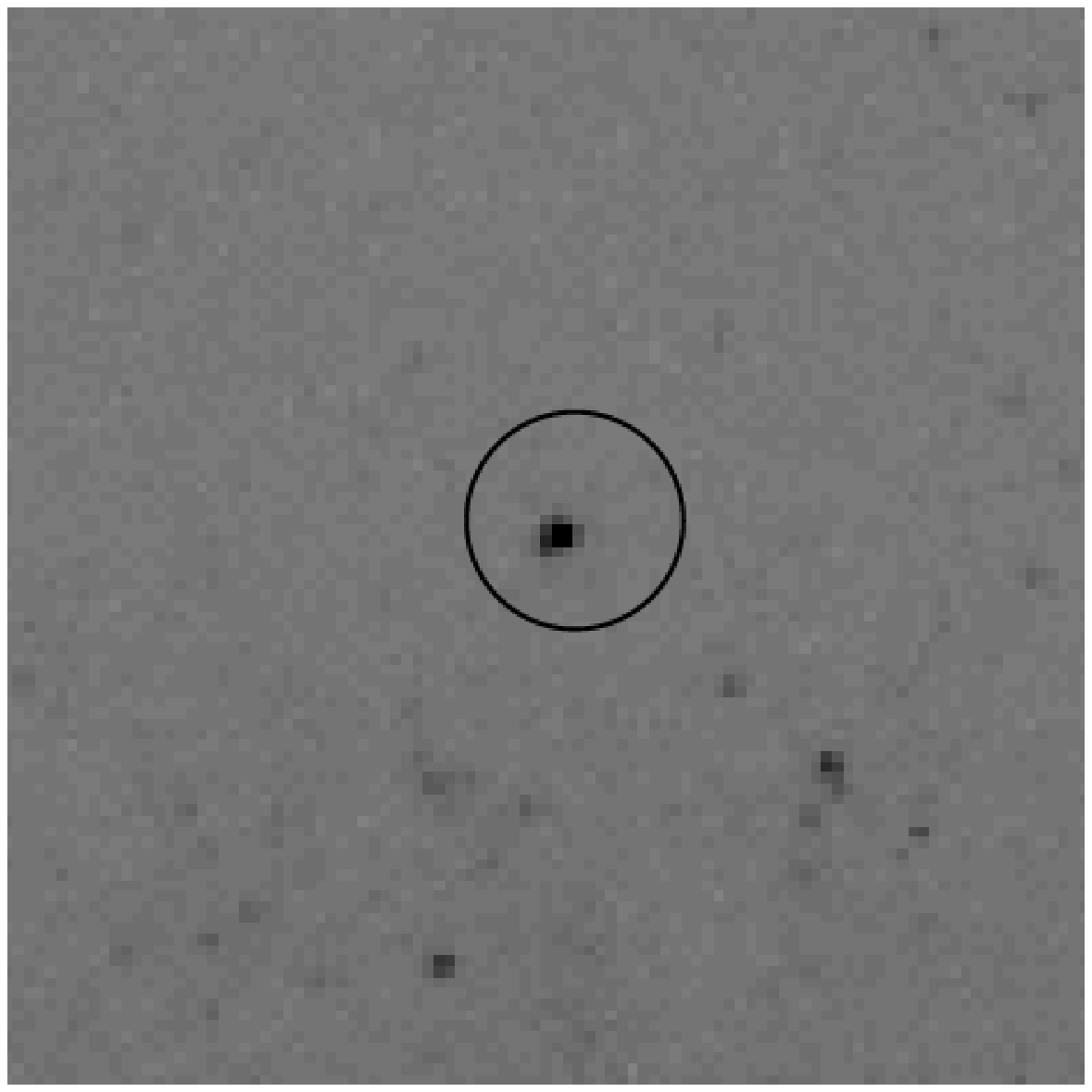}{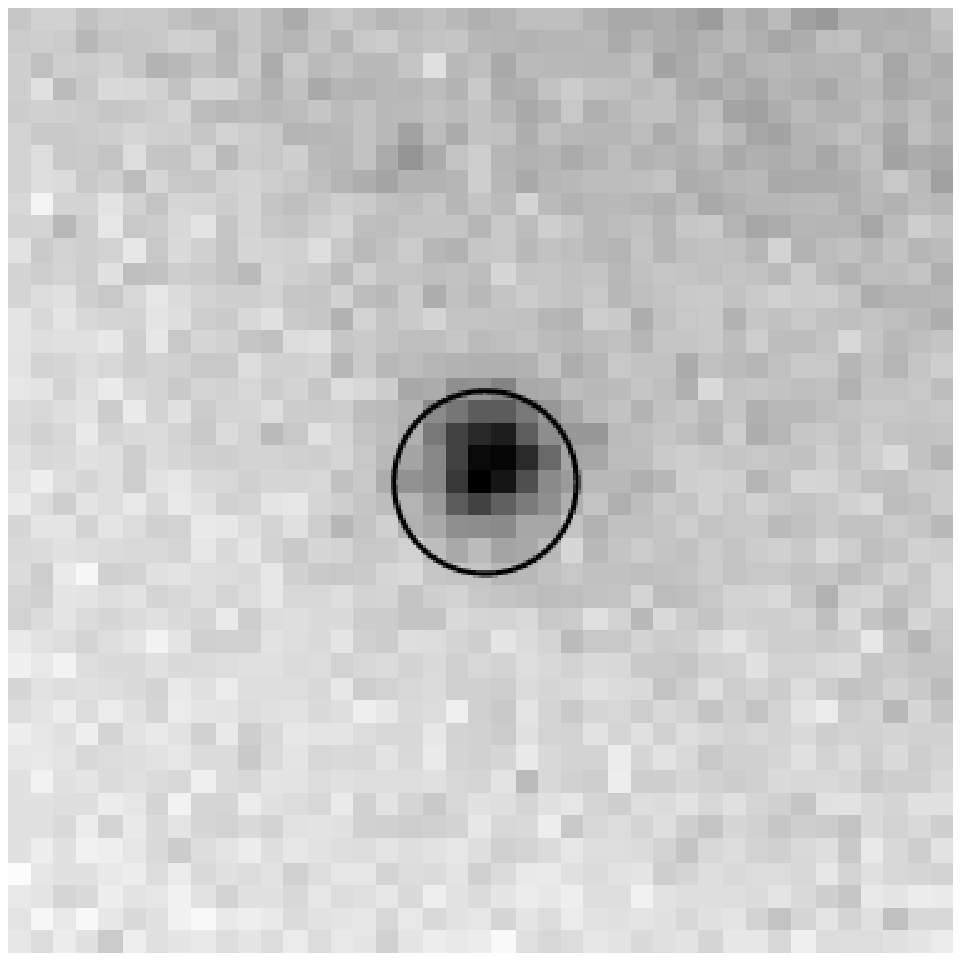}
\caption{$U$ band image (left) and $K_s$ band image (right) showing a 1-arcsecond 
positional error circle centered on the X-ray source, X-37.  Notice the bright 
source well with in this circle.
\label{Fig.1}}
\end{figure}

\clearpage

\begin{figure}
\plotone{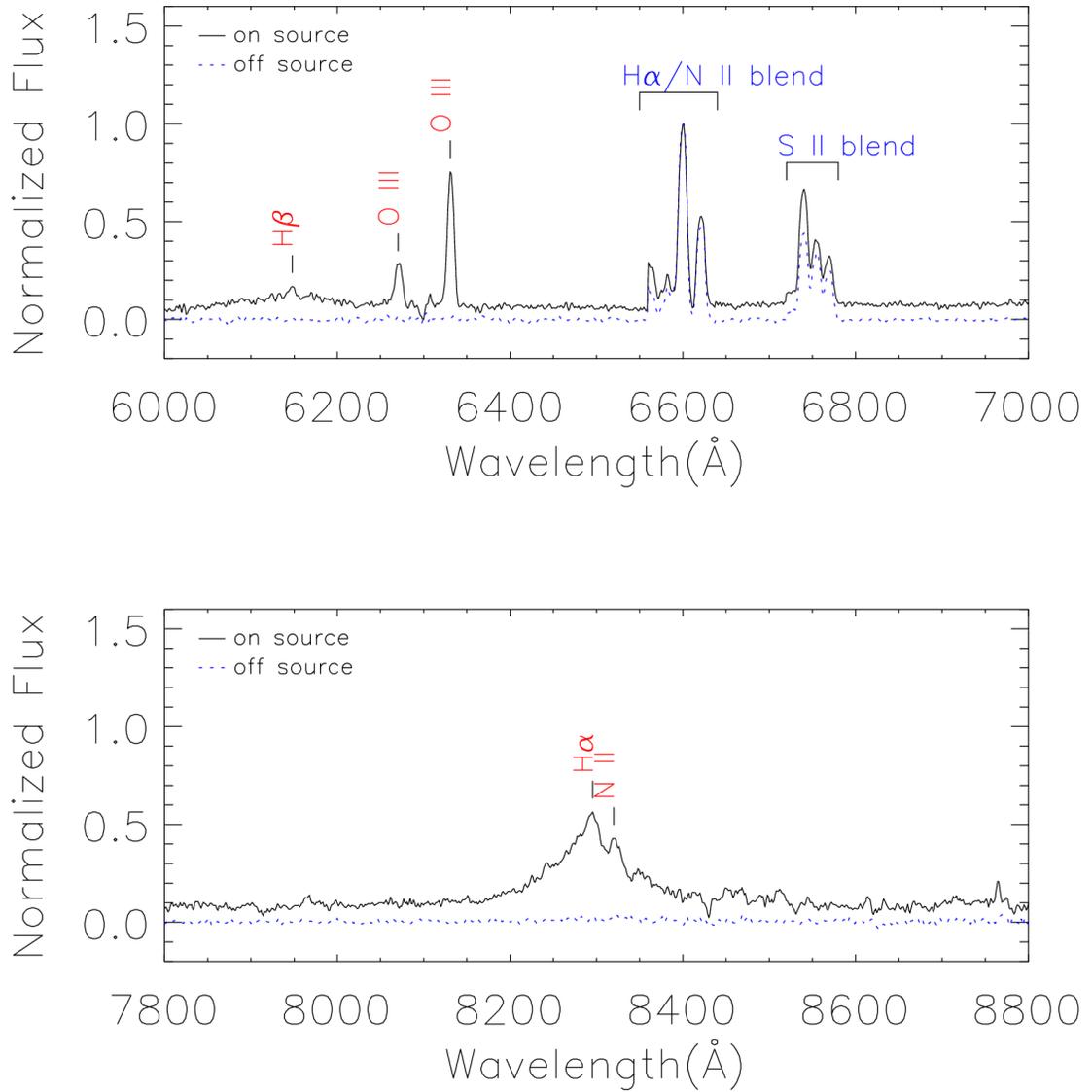}
\caption{X-37 spectrum.  Red text denotes quasar lines, blue text denotes lines
from the Antennae background emission.
\label{Fig.2}}
\end{figure}

\end{document}